\documentclass[%
 aip,
 amsmath,amssymb,
 reprint,%
]{revtex4-1}

\usepackage{graphicx}
\usepackage{dcolumn}
\usepackage{bm}

\usepackage[utf8]{inputenc}
\usepackage[T1]{fontenc}
\usepackage{mathptmx}

\begin{document}

\preprint{AIP/123-QED}

\title[Letter title]{Thermal transport across nanometre gaps:\\ phonon transmission vs air conduction}

\author{A. Alkurdi}
\affiliation{Institut Lumi\`{e}re Mati\`{e}re, UMR5306 Universit\'{e} Claude Bernard Lyon 1-CNRS, Universit\'{e} de Lyon 69622 Villeurbanne Cedex, France}

\author{C. Adessi}
\affiliation{Institut Lumi\`{e}re Mati\`{e}re, UMR5306 Universit\'{e} Claude Bernard Lyon 1-CNRS, Universit\'{e} de Lyon 69622 Villeurbanne Cedex, France}
%

\author{F. Tabatabaei}
\affiliation{Institut Lumi\`{e}re Mati\`{e}re, UMR5306 Universit\'{e} Claude Bernard Lyon 1-CNRS, Universit\'{e} de Lyon 69622 Villeurbanne Cedex, France}
%

\author{S. Li}
\affiliation{Universit\'{e} de Lorraine, CNRS-LEMTA, Nancy, 54000, France} 
%

\author{K. Termentzidis}
\affiliation{CETHIL CNRS-UMR 5008, INSA Lyon, Villeurbanne, 69100, France} 
%

\author{S. Merabia}
\affiliation{Institut Lumi\`{e}re Mati\`{e}re, UMR5306 Universit\'{e} Claude Bernard Lyon 1-CNRS, Universit\'{e} de Lyon 69622 Villeurbanne Cedex, France} 
 \email{samy.merabia@univ-lyon1.fr}

\date{\today}

\begin{abstract}
  Heat transfer between two surfaces separated by a nanometre gap is important for a number of applications ranging from spaced head disk systems, scanning thermal microscopy and thermal transport in aerogels. At these separation distances, near field radiative heat transfer competes with heat transfer mediated by phonons.
  Here we quantity the contribution of phonon assisted heat transfer between apolar solids using lattice dynamics combined with ab-initio calculations. We clearly demonstrate that phonons dominate heat transfer for subnanometre gaps. Strikingly, we conclude that even in the situation where the gap is filled with air molecules, phonons provide the dominant energy channel between the two solids nearly in contact. Our results predict orders of magnitude enhanced phonon heat transfer compared to previous works and bring forward a methodology to analyse phonon transmission across nanoscale vacuum gaps between apolar
  materials.
\end{abstract}

\maketitle

%


Heat may be transferred by three main mechanisms namely conduction, radiation and convection~\cite{Incropera2002}. Convection occurs when heat is transported by the macroscopic motion of a supporting medium.
Heat conduction is also mediated by a supporting medium, and usually not relevant in vacuum~:~it is of common sense that two bodies in vacuum exchange heat only through radiation. The situation becomes less clear however at the nanoscale, when the separation distance between two bodies is in the nanometre range. At these length scales phonons may be transmitted across the gap between the two solids and compete with radiative heat transfer thus contributing in the heat exchange between the two bodies.

Recent experimental investigations have reported measurements of heat transfer in the extreme near field regime, where the separation distance between two objects is in the nanometer range~\cite{Cui2017,Kloppstech2017}. While one study shown moderate deviations to near field predictions (Rytov's theory)~\cite{Rytov1989}, the other reported orders of magnitude  enhancement~\cite{Kloppstech2017}. In this Letter, we quantify the phonon contribution to heat transfer in the ultra near field regime on the basis of ab-initio based lattice dynamics calculations. This method enables us to accurately account for phonon dispersion in the two materials nearly in contact. We demonstrate that at the nanometre scale, phonons are the main energy carriers flowing across the gap, wether it is filled or not with air molecules.  

Heat transfer between two bodies separated by small distances has attracted the attention of physicists for decades~\cite{Cahill2014,Cravalho1967,Hargreaves1969,Domoto1970}. Early studies were motivated by the possibility to observe deviations to Planck's theory of radiation. These pioneering measurements were performed at low temperatures, when Wien's thermal wavelength is macroscopic facilitating the observation of deviations at relative large separation distances~\cite{Kralik2012}. The last decade witnessed intensive effort toward the evidence of extraordinary heat flux in the near field regime~\cite{Kittel2005,Rousseau2009,Shen2009,Narayanaswamy2008a,Altfeder2010,Ottens2011,Lim2015,Kim2015,Song2016,St-Gelais2016}, ever pushing the limits of separation distances.

These fundamental investigations have been fostered by the technological side of the problem. Radiative heat transfer across microscale gaps is at the core of near field thermophotovoltaics, an emerging field of investigation~\cite{Fiorino2018}. At shorter separation distances, the applications concern thermal rectifiers~\cite{Otey2010}, heat assisted magnetic recording~\cite{Wu2016} and scanning thermal microscope~\cite{Gomes2015}. Thermal transport across nanocavities controls also the thermal conductivity of silica aerogels~\cite{Zeng1994}.

Despite its technological importance, thermal transport between two bodies nearly in contact still resists a full theoretical treatment. One of the yet unsolved question is how to describe the gradual transition between radiative and conductive heat transfer in the extreme near field regime. Depending on the separation distance, different regimes may be distinguished. Indeed, when two bodies are separated by macroscopic distances, thermal transport is very well described by Planck's blackbody radiation law~\cite{Planck1914,Incropera2002}. When the separation distance becomes comparable to Wien's thermal wavelength ($\lambda \simeq 10 \mu m$ at $300$ K), radiation can exceed by orders of magnitude blackbody law. In this regime, photons emitted by one of the two media tunnel across the gap, and the overlapping evanescent waves yield enhanced heat transfer. At the other extreme when two bodies are in contact, heat transfer is mediated by phonon transmission. At nanometre distances, the distinction between radiation and conduction becomes somewhat blurred, and this is precisely this regime that we investigate in this Letter.  

Several mechanisms have been invoked to come into play when the separation distance is ultra small~\cite{Chiloyan2015,Prunnila2010,Ezzahri2014,Xiong2014,Messina2018}. They have in common to involve the coupling between phonons and electric fields, a situation relevant to piezoelectric or polar materials. Here, we consider phonon contribution to heat transfer across {\it apolar} materials. We concentrate on this type of materials, as there is no coupling between phonons and electric fields. Our calculations show that phonons are the main energy channel for nanometre gaps, thus correcting by orders of magnitude previous estimates of phonon assisted thermal transport~\cite{Pendry2016}.

\begin{figure}[h]
\centering
\includegraphics[scale=0.20]{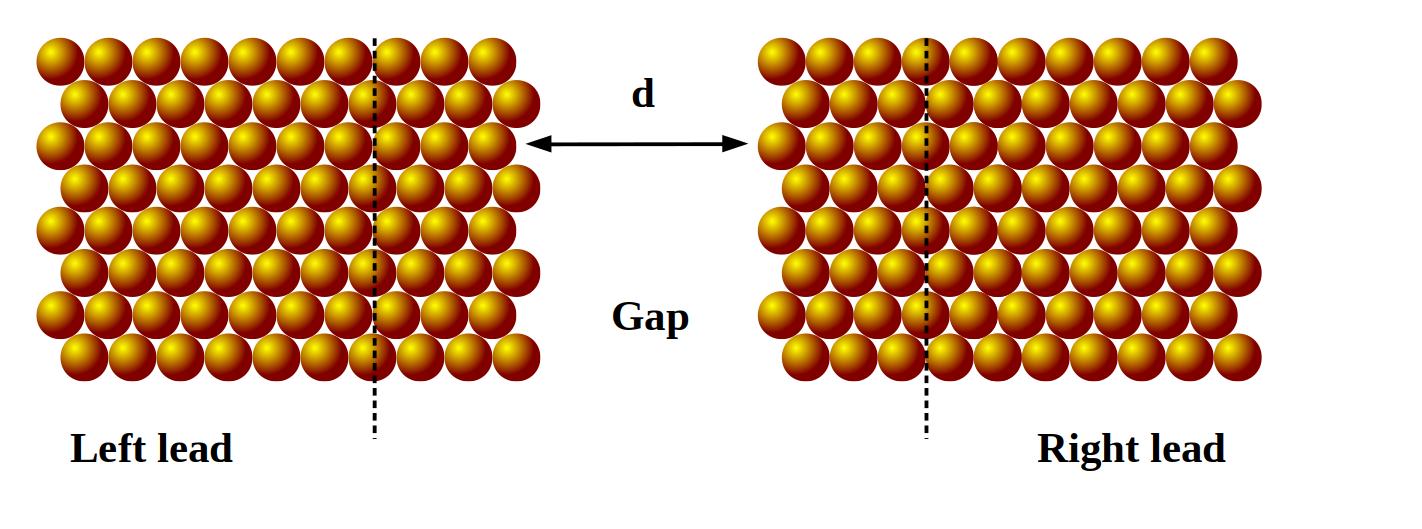}
\caption{\label{fig:illustration}(Color online) Illustration of the configuration considered~:~two semi-infinite solids separated by a gap of width $d$. }
\end{figure}

To estimate the contribution due to phonon across nanoscale gaps, we employ ab-initio based lattice dynamics (LD) calculations~\cite{Alkurdi2017a,Alkurdi2017b}. The principle of the method is inspired by Zhao and Freund work~\cite{Zhao2005}, which considered empirical potentials to describe the atomic lattice dynamics at the interface between two solids. We basically extend Zhao and Freund's method to account for harmonic force constants obtained from ab-initio first principles calculations. In brief, the system is divided in three regions namely, the left and the right leads and the gap junction, as illustrated in fig.~\ref{fig:illustration}.

In the semi-infinite bulk leads, we solve the dynamical equations:
\begin{equation}
  (\phi_{\alpha,\beta}-m \omega^2 \delta_{\alpha,\beta}) u_{\beta}(\omega) = 0
\end{equation}  
where $\Phi_{\alpha,\beta}$ denotes the harmonic force constant tensor, $\omega$ is the phonon frequency and $u_{\beta}(\omega)$ denotes the components of the atomic displacement
along the direction $\beta$.

In the junction region, the equations to be solved write~:
\begin{eqnarray}
  (\phi^{AA}_{\alpha,\beta}-m_A \omega^2 \delta_{\alpha,\beta}) u^A_{\beta}(\omega) +  \phi^{AB}_{\alpha,\beta}  u^B_{\beta}(\omega)  = 0 \\
    \phi^{BA}_{\alpha,\beta}  u^A_{\beta}(\omega) + (\phi^{BB}_{\alpha,\beta}-m_B \omega^2 \delta_{\alpha,\beta}) u^B_{\beta}(\omega) = 0 
\end{eqnarray}  
here the subscripts $A,B$ denote the two materials, and we have introduced the corresponding Hessian matrices $\phi^{AA}, \phi^{BB}, \phi^{AB}, \phi^{BA}$ with obvious notations.
The displacement $u^A$ in the incoming medium may be split in an incoming wave having a wavevector $k_i$ and $n_r$ reflected waves, each having a wavevector $k_r$. Similarly, the displacement $u^B$ in the   
other medium is a sum of $n_t$ waves characterized by their wavevectors $k_t$. The numbers $n_t,n_r$ and the expressions of the wavevectors $k_r,k_t$ are given by the conservation equations at the interface,
see \cite{Zhao2005,Alkurdi2017b} for further details. From the knowledge of the transmitted displacement, one can infer the {\em mode} dependent transmission coefficient and interface conductance~:

\begin{equation}
  \mathcal{T}(\mathbf{k}_t,\nu)= \frac{\rho_A}{\rho_B}\frac{v^{z}_{g,t}.\vert A_{t}\vert^2}{v^{z}_{g,i}.\vert A_{i}\vert^2} \label{eq:tdef}
\end{equation}
where $\nu$ is an index denoting the phonon branch index, $\rho_A$ and $\rho_B$ are the mass density characterizing the two materials respectively, $v^{z}_{g,t}$ is the group velocities of the transmitted phonons projected along the direction perpendicular to the interface, $A_{t}$ and $A_i$ are the amplitudes of the transmitted and incident wave respectively. The mode-dependent thermal conductance has the expression~:
\begin{equation}
G(\mathbf{k}_t,\nu) = \frac{1}{V}  v^{z}_{g,t} \hbar \omega \frac{\partial f(\omega,T)}{\partial T} \mathcal{T}(\mathbf{k}_t,\nu)
\end{equation}
where $V$ is the volume of the system and $f$ denotes the phonon occupation density.


For gold, the bulk harmonic force constants are calculated with the code VASP~\cite{Kresse1996a,Kresse1996b} using the Generalized Gradient Approximation (GGA-PBE)~\cite{Perdew1996,Perdew1997} . 
For silicon, we extracted the values from the literature~\cite{Aouissi2006}.
In the S1 of the Supplementary material~\cite{supplementary_material}, we illustrate the good agreement between the calculated bulk phonon spectra and available experimental data.

For the interaction between the materials across the gap, we consider a Lennard-Jones potential~\cite{Israelachvili}:
\begin{equation}
\Phi_{zz}(z) = 8 \pi \rho \epsilon \sigma \left( \left(\frac{\sigma}{z} \right)^{11} - \left(\frac{\sigma}{z} \right)^5 \right)
\label{Lennard_Jones}
\end{equation}
where $\rho$ is the material density and $z$ is the distance to the other medium. 
The Lennard-Jones parameters $\epsilon,\sigma$ are extracted from~\cite{Heinz2008} for metals, and for semiconductors we used the values derived from the knowledge of the material Hamaker's constant~\cite{Israelachvili}, as listed in S2 of~\cite{supplementary_material}. 
The atomistic calculations will be compared with the predictions of a generalized acoustic mismatch model~(AMM), describing phonon transport across the solid/vacuum gap/solid interface.
In this model, the transmission coefficient is supposed to be given by $\mathcal{T}(\omega)=1/(1+(\omega/2K)^2 z_m^2)$ where $z_m=\rho_{m}c$ is the acoustic impedance of the material with $\rho_m$ and $c$ the mass density and speed of sound respectively, and $K=n_s (\frac{d^2 V}{d z^2})_{z=d}$ is the interface compliance, with $n_S$ the density of surface atoms and $V(z)=  8 \pi \rho^2 \epsilon \sigma^4 \left( \frac{1}{720} \left( \frac{\sigma}{z} \right)^{8} -  \frac{1}{24} \left(\frac{\sigma}{z} \right)^2 \right)$ is the {\em total} interaction potential per unit of surface between the two solids. Further details may be found in the S3~\cite{supplementary_material}. 
For gold, we will compare the relative contribution of phonons and electrons to heat transfer. To estimate the importance of this latter energy channel, we will employ an analytical expression recently derived~\cite{Messina2018}, which describes electronic heat transfer as a tunneling effect. Here, we will consider a zero bias voltage situation to be consistent with the atomistic calculations. 
Last, the contribution due to the presence of air molecules in the gap is estimated following Devienne~\cite{Devienne1965}, as detailed in S4 of~\cite{supplementary_material}. To assess the relevance of Devienne's expression to hold at the nanoscale, we have performed parallel molecular dynamics simulations taking into account the presence of gas molecules in the gap between two gold surfaces. The value obtained for the thermal conductance was found to be smaller than Devienne's prediction. Therefore, the values of the air conduction conductance that we will report here should be considered as an upper bound of the contribution of molecules in the gap.


\begin{figure}[h]
\centering
\includegraphics[scale=0.3]{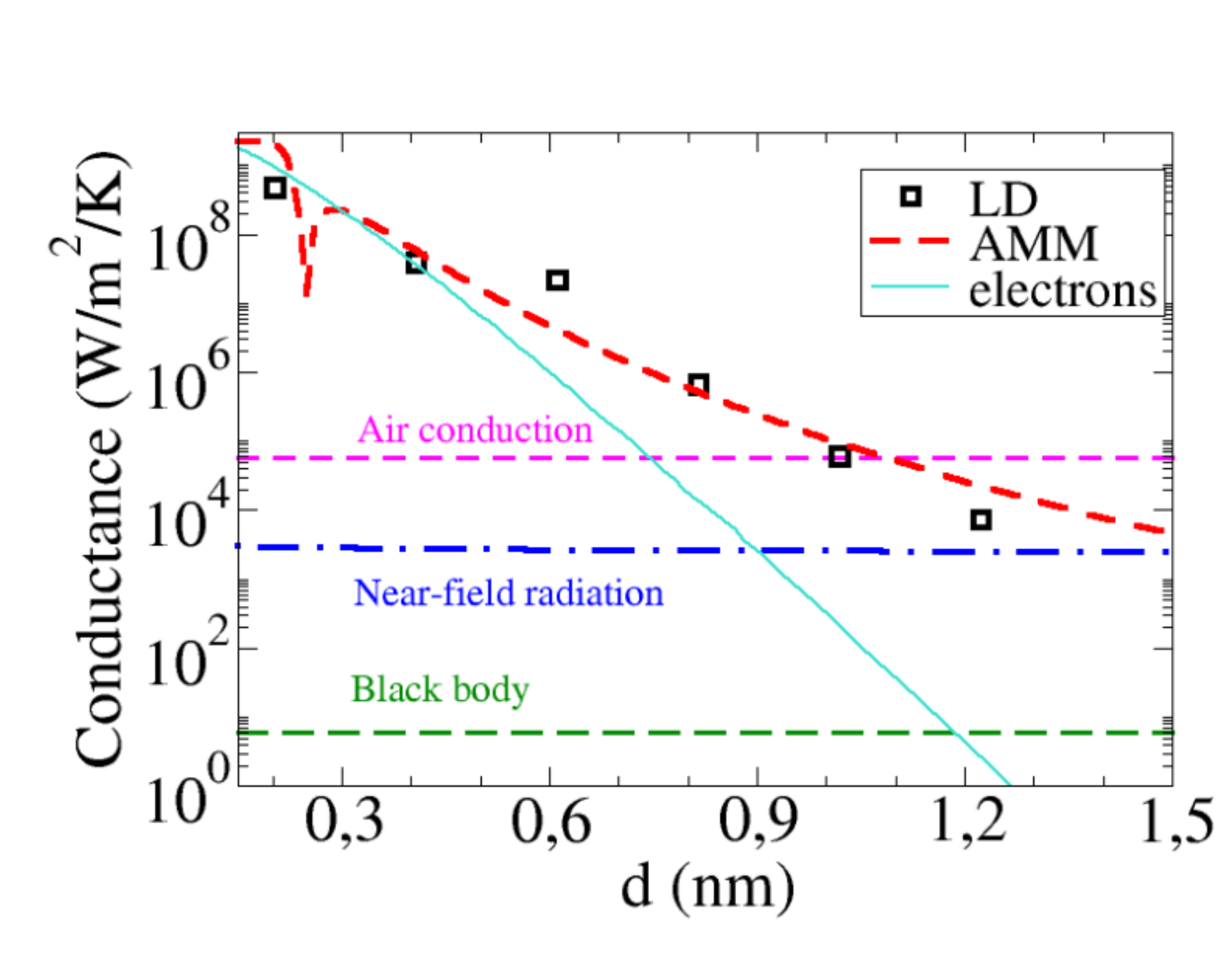}
\includegraphics[scale=0.3]{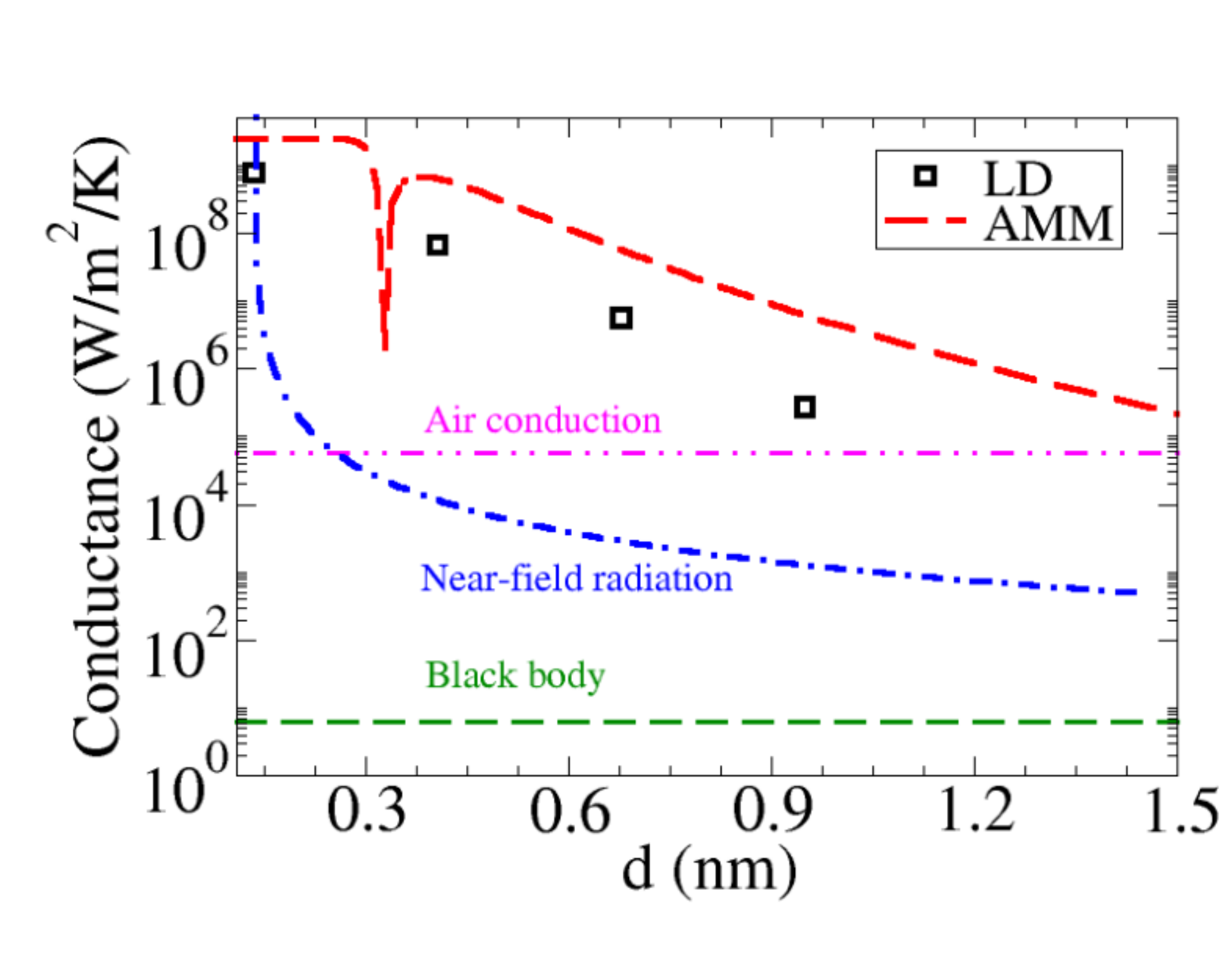}
\caption{\label{fig:G_gap}(Color online) Top~:~Thermal gap conductance between two gold solids separated by a nanometre gap of width $d$, calculated at room temperature.  The continuous red line is the generalized AMM model. The blue line is the electronic heat transfer calculated from~\cite{Messina2018}. Near-field radiative predictions are taken from \cite{Chapuis2008}. Bottom~:~Thermal boundary conductance between two silicon solids separated by a nanometric gap of width $d$. Near-field radiative prediction is taken from \cite{Sellan2012}.   }
\end{figure}

Figure~\ref{fig:G_gap} presents the atomistic calculations results for the thermal transport across the gap between two gold solids. The calculated conductance due to phonons is compared to the contribution due to different energy channels.  Clearly, the contribution due to phonons is by far the largest, for the range of gap widths analyzed here. It is orders of magnitude higher than black body conductance
$G_{BB} \simeq 6$ W/m$^2$/K~\cite{Incropera2002}. Less expectedly, phonon heat transfer beats near field radiative heat transfer also by orders of magnitude.
Also, we remark that electronic heat transfer is smaller than phonon tunelling for gap thicknesslarger than $0.4$ nm. Last, the contribution due to phonons is higher than air conduction for gap widths smaller than $1$ nm. This implies that if the gap is filled with air molecules, long range interactions between the two solids contribute much more to the heat transfer than ballistic transport mediated by the molecules. It is remarkable that the atomistic calculations yield conductance levels comparable to the AMM predictions. We will rationalize this observation through the analysis of the spectral phonon transmission across the gap. Note that the dip present in the AMM conductance for small gaps is related to the non monotonicity of the Lennard-Jones spring constant, driven by short-range interactions. 

Figure~\ref{fig:G_gap} demonstrates that the same qualitative conclusions may be drawn for the silicon/gap/silicon system. In this latter case however, the atomistic calculations give conductance levels lower than the AMM predictions. Nevertheless, phonon contribution still dominates air conduction below $1$ nm.

\begin{figure}[h]
\centering
\includegraphics[scale=0.19]{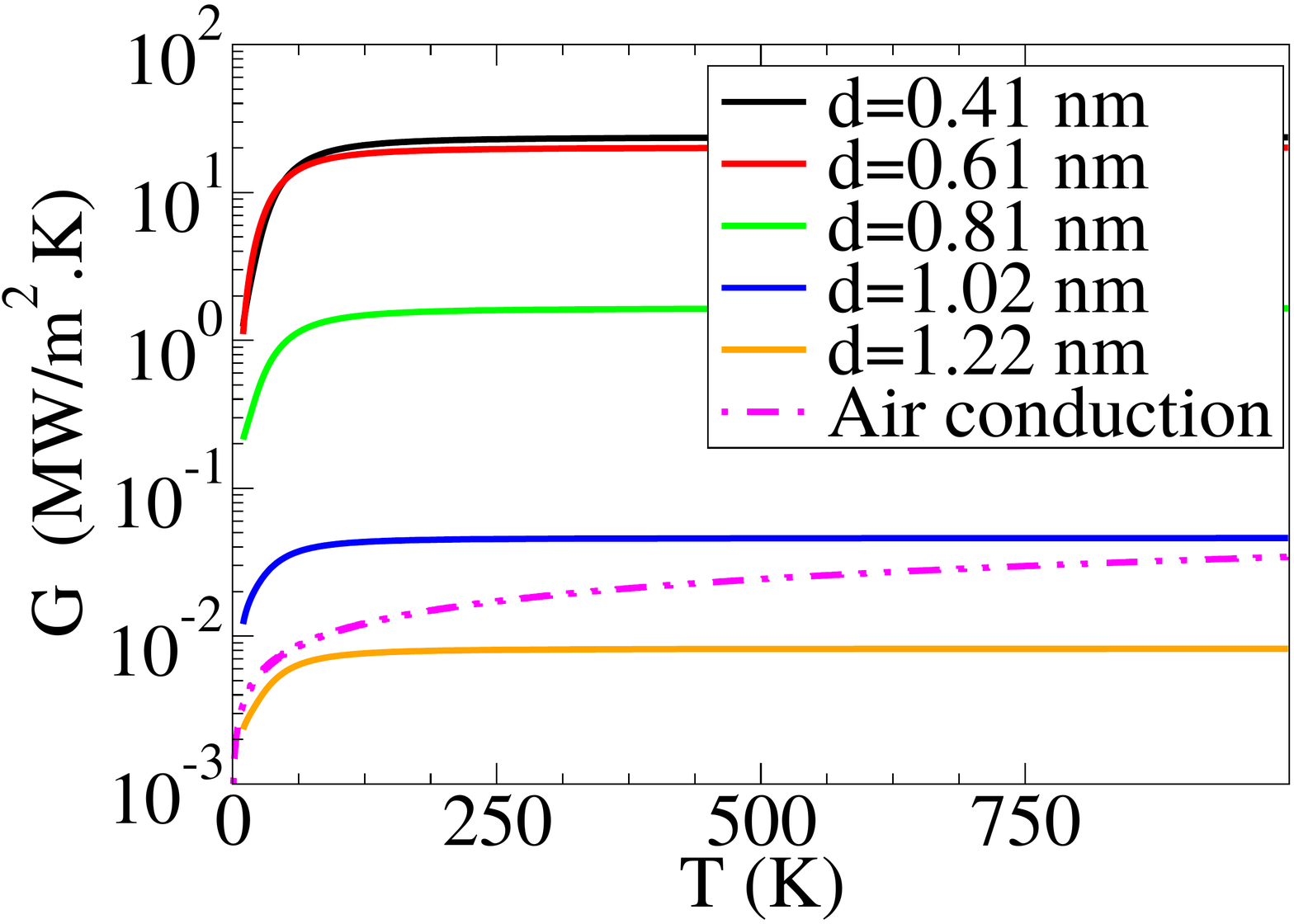}
\includegraphics[scale=0.19]{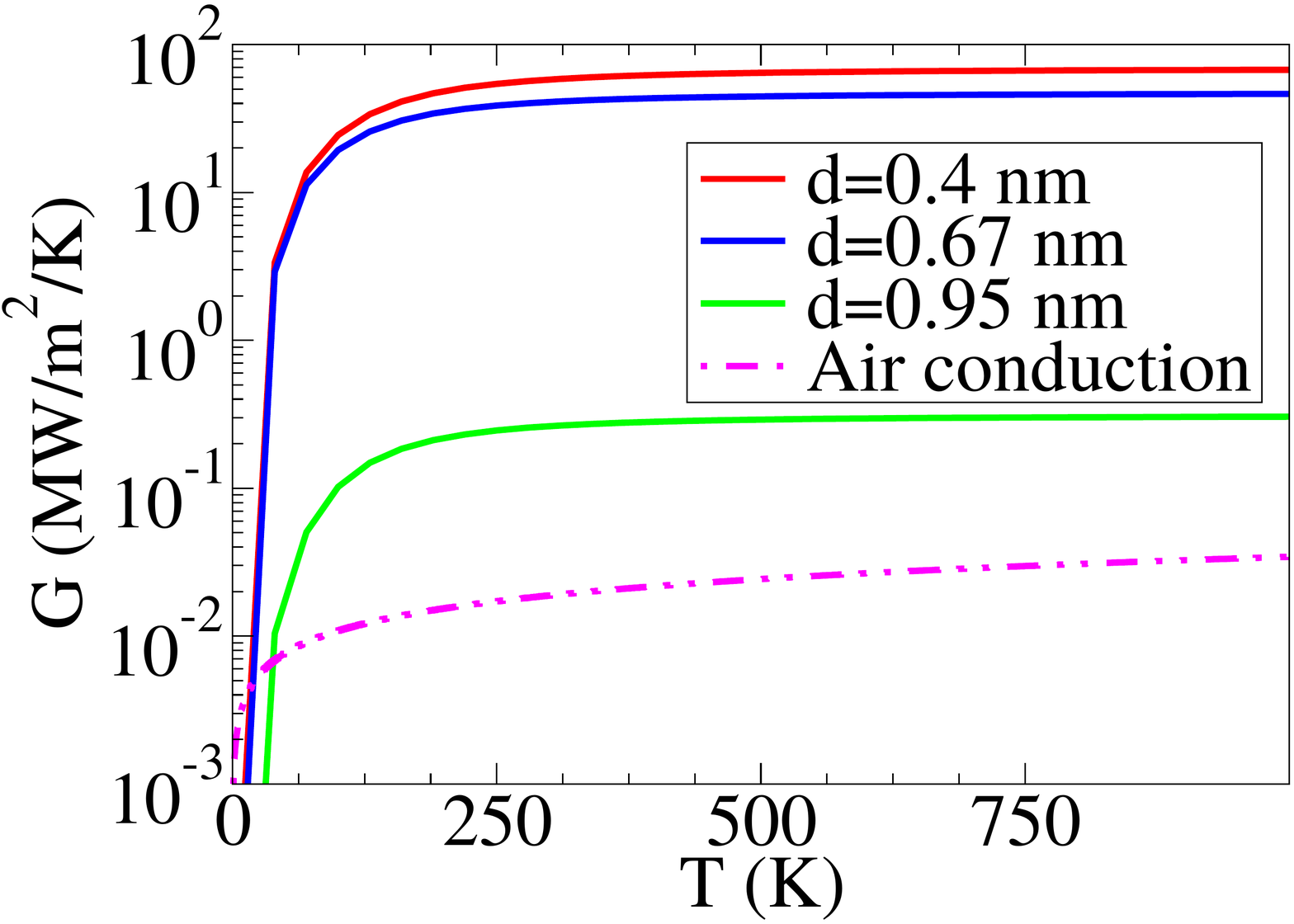}
\caption{\label{fig:G_gap_Temperature}(Color online) Thermal boundary conductance between two gold solids~(top) and two silicon solids~(bottom) separated by a nanometre gap $d$ as a function of temperature.
The contribution due to air conduction is shown with dashed purple lines.}

\end{figure}

The dominance of phonon over air conduction was shown here at room temperature. To appraise the generality of this behavior, the temperature dependence of the conductance characterizing the two main energy channels-phonon transmission and air conduction-has been reported in fig.~\ref{fig:G_gap_Temperature}. Clearly over the wide range of temperatures considered, the contribution due to phonons is the highest. Only at cryogenic temperatures, a very limited number of acoustic modes is populated, and air conduction may become the leading channel. 

\begin{figure}[h]
\centering
\includegraphics[scale=0.19]{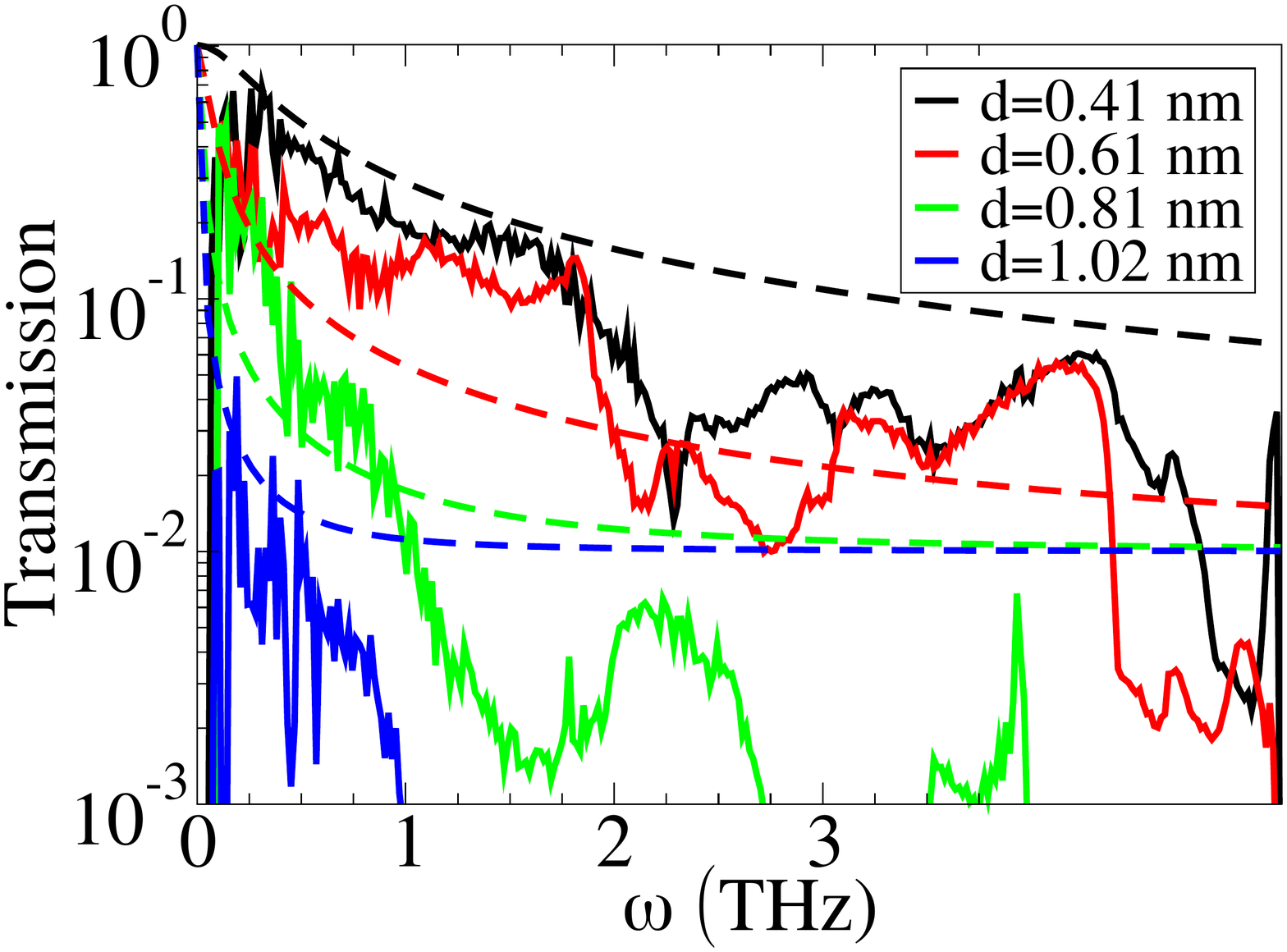}
\includegraphics[scale=0.25]{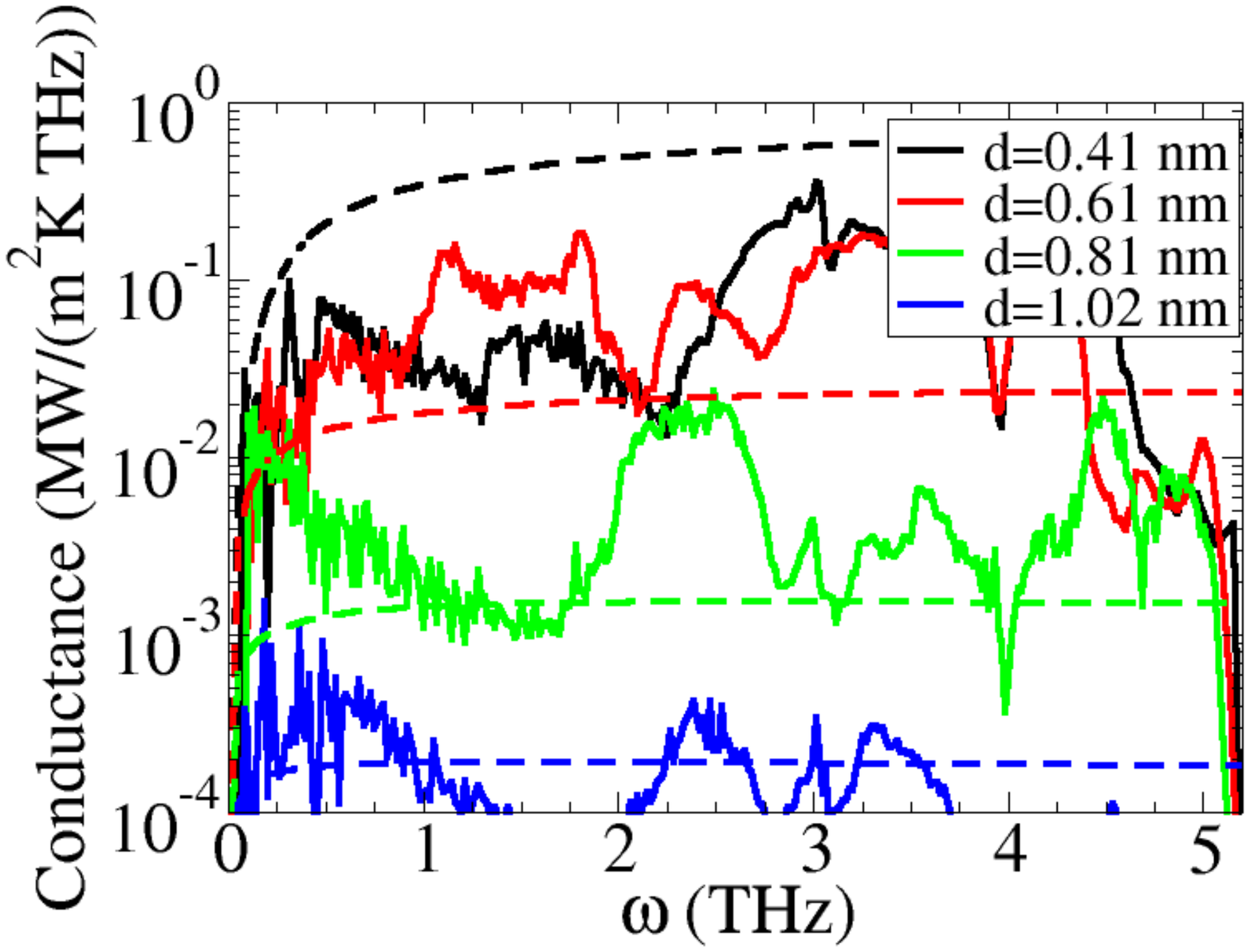}
\caption{\label{fig:T_w_gap_Au}(Color online) Phonon transmission coefficient and spectral conductance between two gold solids separated by a nanometre gap $d$, as a function of frequency.
 The solid lines display the atomistic calculations while the dashed lines are the AMM predictions.}
\end{figure}

To get insight in the microscopic mechanisms behind thermal conduction in the gap, we discuss now the spectral contribution to the interfacial heat transfer mediated by phonons.
Of particular interest, we analyze how acoustic models can accurately capture the contribution of heat flow due to phonons. 
We concentrate first on the case of gold, for which the spectral transmission and conductance are plotted in fig.~\ref{fig:T_w_gap_Au}. For all the gap widths analyzed, only low frequency phonons are transmitted across the gap. This behavior is well captured by AMM models at low frequency, while these latter models tend to overestimate slightly phonon transmission for higher frequencies. On the opposite, the AMM predictions underestimate somewhat the conductance, except for the narrowest gap analysed. This conclusion is in full line with the total conductance calculated and presented in fig.~\ref{fig:G_gap}. The  relatively low values of the conductance predicted by AMM may be explained by the underestimation of the density of intermediate frequencies, contributing to undervalue the spectral conductance.

\begin{figure}[h]
\centering
\includegraphics[scale=0.19]{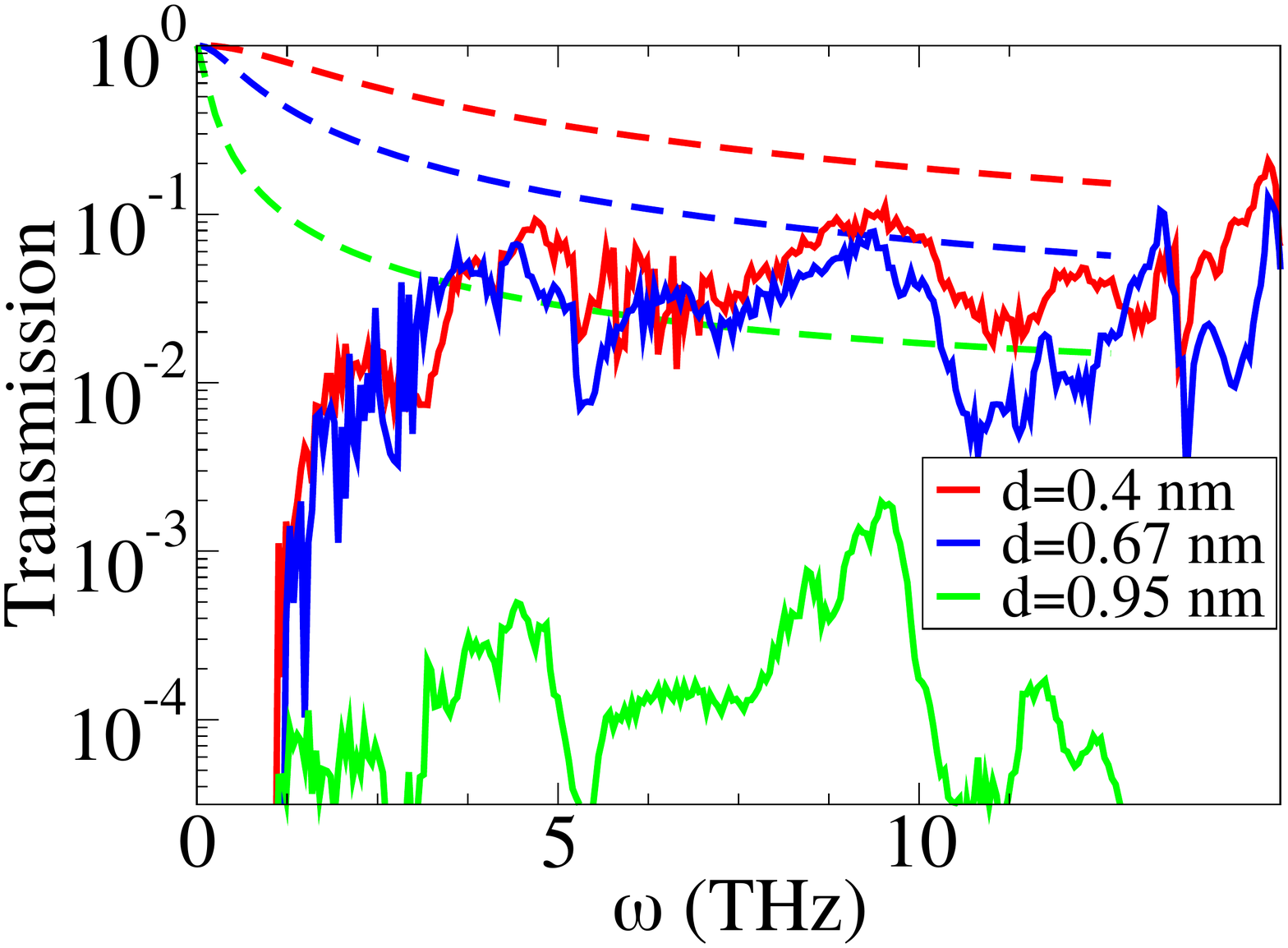}
\includegraphics[scale=0.19]{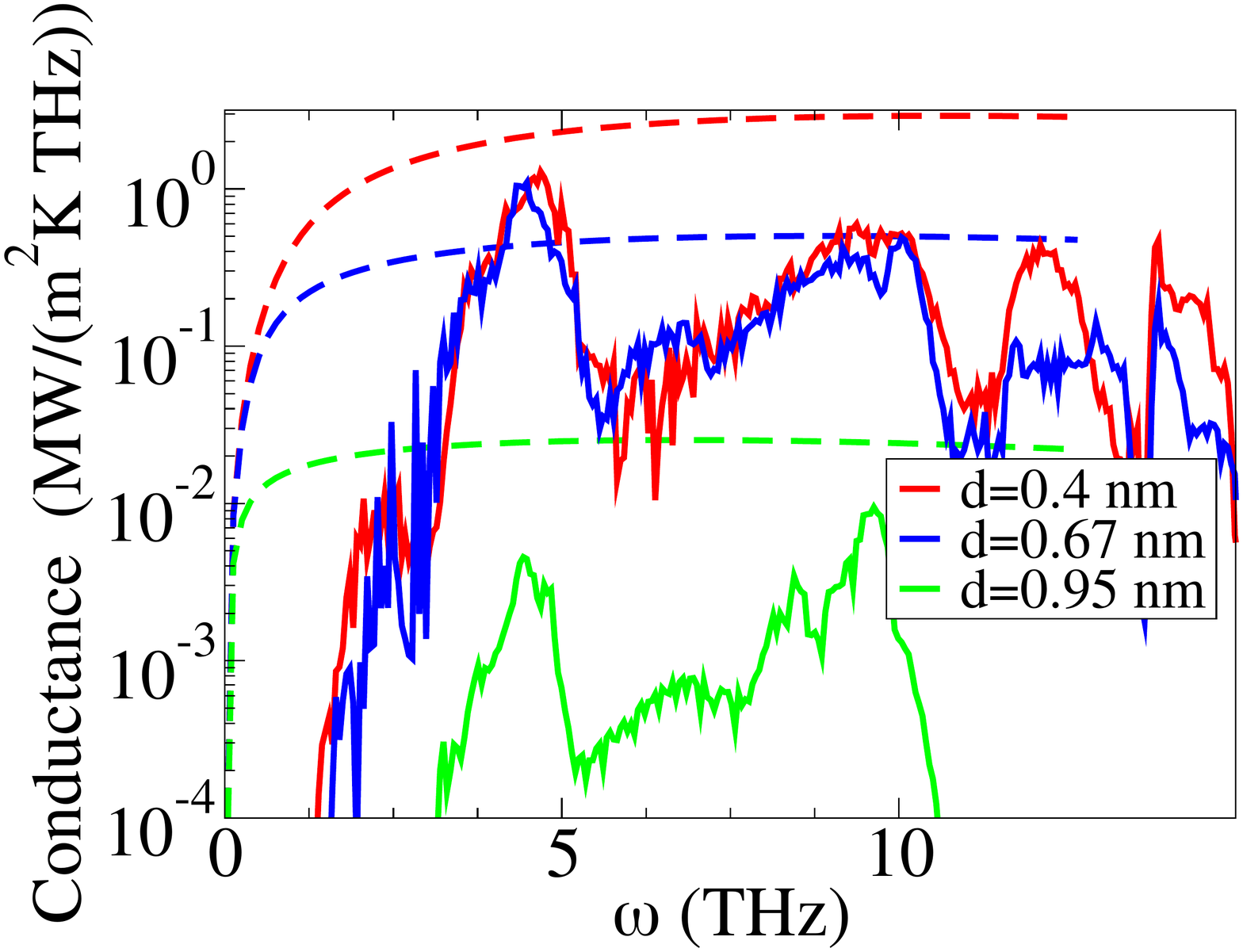}
\caption{\label{fig:T_w_SiSi_gap_vdW}(Color online) Same as fig.~\ref{fig:T_w_gap_Au} for Silicon.}
\end{figure}

We now turn to the case of silicon, as illustrated in fig.~\ref{fig:T_w_SiSi_gap_vdW}. Here a different scenario emerges~:~the transmission coefficient of long wavelength phonons is low, and only intermediate frequencies phonons can tunnel across the gap. But even the transmission coefficient of these modes is overestimated by AMM, leading to a discrepancy in the spectral conductance. 
This analysis explains the relatively low conductance of silicon interfaces, as displayed in fig.~\ref{fig:G_gap}.  

Our calculations involve so far phonon transmission, and it is legitimate to wonder wether alternative energy channels may contribute to heat transfer across the gap. As shown in fig.~\ref{fig:G_gap} and in the experiments of Kloppstech et {\em al.}~\cite{Kloppstech2017}, electron tunneling contributes to heat transfer when the gap distance becomes smaller than $0.4$ nm, thus precluding any direct electron-electron scattering mechanism for larger separation distances. However, metal electrons may play an indirect role in heat transfer through their couplings with phonons. Indeed, electrons couple with phonons in different ways, as summarized in~\cite{Lombard2015}.  Of particular interest here, is the situation of weak electron-electron coupling through the gap. Under these conditions, the electronic temperature should obey an adiabatic condition at the boundaries. In the Section S5 of~\cite{supplementary_material}, we show that the effect of the electron-phonon coupling is to reduce the effective thermal boundary conductance by an amount~:~$G_{\rm eff} \simeq G/(1+2G/\sqrt{k_p G_{\rm ep}})$, where $G$ is the phonon-phonon conductance that we already calculated, $k_p$ is the phononic contribution to the metal conductivity and $G_{\rm ep}$ is the electron-phonon coupling. This latter expression generalizes the expression derived in~\cite{Reddy2004} in the case of metal/semiconductor interface. Considering the values of $k_p$ and $G_{\rm ep}$ for gold~\cite{Jain2016,Lin2008}  $k_p=2$ W/mK; $G_{\rm ep}=2.5 \; 10^{16}$ W.m$^{-3}$.K$^{-1}$, the resulting values of the conductance is $0.11$ MW.m$^{-2}$.K$^{-1}$, which is higher than the conductance due to the ballistic motion of air molecules. Note that gold has a relatively low electron-phonon coupling constant, so that the electron-phonon interfacial conductance is even higher for other metals~($0.62$ MW.m$^{-1}$.K$^{-1}$ for aluminium). Interfacial electron-phonon transfer will therefore not impede interstitial heat transfer below air conduction level.   
In summary, we investigated phonon mediated thermal transport in the extreme near field regime using lattice dynamics calculations. We demonstrated that for nanometre gaps, phonons are the main energy carriers exceeding by orders of magnitude near field radiative heat transfer. Strikingly, the contribution of phonons exceeds also largely ballistic transport assisted by gas molecules in the gap.
This implies that even in the situation where the gap is filled with air, long range van der Waals interactions provide the dominant heat transfer mechanism. Our approach can be generalized to any apolar material, opening the door to accurate ab-initio based computations of phonon mediated near feld heat transfer, as well as studied experimental situations~\cite{Cui2017,Kloppstech2017}.

\begin{acknowledgements}
The authors acknowledge fruitful discussions with P.-O. Chapuis, T. Albaret, A.~Ayari, P.~Vincent, R. Messina and P. Ben Abdallah. Financial support from Labex projet iMust ATTSEM and H2020 programme FET-open project EFINED (number 756853) is acknowledged.
\end{acknowledgements}

\nocite{*}

\end{document}